\documentclass{emulateapj}
\usepackage{graphicx,natbib,amsmath} 
\usepackage{CJKutf8}

\newcommand{\Lsun}{$L_{\odot}$}
\newcommand{\Msun}{$M_{\odot}$}
\newcommand{\LIR}{$L_{\rm IR}$}  
\newcommand{\mum}{$\mu m$}

\newcommand{\Ha}{H$\alpha$}
\newcommand{\NII}{[N\,{\sc ii}\rm]}

\begin{document}

\title{Kinematic classifications of local interacting galaxies: implications for the merger/disk classifications at high$-z$}

\author{Chao-Ling Hung \begin{CJK*}{UTF8}{bsmi}(洪肇伶)\end{CJK*}\altaffilmark{1,2}}
\author{Jeffrey A. Rich\altaffilmark{3,4}}
\author{Tiantian Yuan\altaffilmark{5}}
\author{Kirsten L. Larson\altaffilmark{1}}
\author{Caitlin M. Casey\altaffilmark{6}}
\author{Howard A. Smith\altaffilmark{2}}
\author{D. B. Sanders\altaffilmark{1}}
\author{Lisa J. Kewley\altaffilmark{5}}
\author{Christopher C. Hayward\altaffilmark{7,8}}

\affil{\altaffilmark{1} Institute for Astronomy, University of Hawaii, 2680 Woodlawn Drive, Honolulu, HI 96822, USA; clhung@ifa.hawaii.edu}
\affil{\altaffilmark{2} Harvard-Smithsonian Center for Astrophysics, Cambridge, MA 02138, USA}
%\affil{\altaffilmark{3} SAO Pre-doctoral Fellow}
\affil{\altaffilmark{3} Infrared Processing and Analysis Center, Caltech 100-22, Pasadena, CA 91125, USA}
\affil{\altaffilmark{4} Observatories of the Carnegie Institution of Washington, 813 Santa Barbara Street, Pasadena, CA 91101, USA}
\affil{\altaffilmark{5} Research School of Astronomy and Astrophysics, Australian National University, Cotter Rd., Weston ACT 2611, Australia}
\affil{\altaffilmark{6} Department of Physics and Astronomy, University of California at Irvine, 2162 Frederick Reines Hall, Irvine, CA 92697, USA}
\affil{\altaffilmark{7} Heidelberger Institut f\"{u}r Theoretiche Studien, Schloss-Wolfsbrunnenweg 35, 69118 Heidelberg, Germany}
\affil{\altaffilmark{8} TAPIR, Mailcode 350-17, California Institute of Technology, 1200 E. California Boulevard, Pasadena, CA 91125, USA}

\submitted{Submitted 2014 July 24; accepted 2015 March 16}

\begin{abstract}

The classification of galaxy mergers and isolated disks is key for understanding the relative importance of galaxy interactions and secular evolution during the assembly of galaxies.
The kinematic properties of galaxies as traced by emission lines have been used to suggest the existence of a significant population of high$-z$ star-forming galaxies consistent with isolated rotating disks.
However, recent studies have cautioned that post-coalescence mergers may also display disk-like kinematics.
To further investigate the robustness of merger/disk classifications based on kinematic properties, we carry out a systematic classification of 24 local (U)LIRGs spanning a range of galaxy morphologies: from isolated spiral galaxies, ongoing interacting systems, to fully merged remnants.
We artificially redshift the Wide Field Spectrosgraph (WiFeS) observations of these local (U)LIRGs to $z=1.5$ to make a realistic comparison with observations at high$-z$, and also to ensure that all galaxies have the same spatial sampling of $\sim900$ pc.
Using both kinemetry-based classifications and visual inspection, we find that the reliability of kinematic classification shows a strong trend with the interaction stage of galaxies. 
Mergers with two nuclei and tidal tails have the most distinct kinematic properties compared to isolated disks, whereas a significant population of the interacting disks and merger remnants are indistinguishable from isolated disks based on the measurements of kinematic asymmetries or visual inspection.
The high fraction of late-stage mergers displaying disk-like kinematics reflects the complexity of the dynamics during galaxy interactions.
However, the exact fractions of misidentified disks and mergers depend on the definition of kinematic asymmetries and the merger/disk classification threshold when using kinemetry-based classifications.
Our results suggest that additional merger indicators such as morphological properties traced by stars or molecular gas are required to further constrain the merger/disk classifications at high$-z$.

\end{abstract}


\section{Introduction}

Galaxy interactions play a key role in driving the growth of galaxies and transforming galaxy morphology \citep[e.g.][]{Toomre1972,Toomre1977,White1978,Schweizer1982,Barnes1988,Wright1990,Barnes1992a}.
During the merger processes, gas in disk galaxies loses its angular momentum and falls toward the center, inducing intense starbursts and active galactic nuclei (AGN) activities \citep[e.g.][]{Barnes1996,Mihos1996,Sanders1996}.
The strong feedback from luminous AGNs and QSOs can then disperse the surrounding gas, halt the black hole growth, and quench the star formation \citep[e.g.][]{Sanders1988,Hopkins2006a}.

One essential way to identify interacting galaxies/mergers is through the characterization of galaxies' optical morphological properties \citep[e.g.][]{Surace1998,Farrah2001,Veilleux2002}.
Numerical simulations have shown that violent encounters of gas-rich spiral galaxies can result in various observational footprints of interactions such as bridges and tidal tails \citep[e.g.][]{Toomre1972,Barnes1992}.
While visual classification is still used for identifying mergers out to $z\sim1-3$ \citep[e.g.][]{Bell2005,Dasyra2008,Kartaltepe2010a,Hung2013,Kartaltepe2012}, various automatic classification schemes have also been developed and applied to large extragalactic surveys \citep[e.g.][]{Abraham2003,Conselice2003,Lotz2004,Law2007,Freeman2013}.
Identifying such merger signatures with either visual or automatic classifications at $z\sim1-3$ is challenging due to surface brightness dimming and band-shifting \citep[e.g.][]{Hibbard1997,Petty2009}, which can lead to a significant underestimation of the occurrence of mergers \citep[e.g.][]{Abraham1996,Overzier2010,Hung2014,Petty2014}. 
Meanwhile, $z\sim1-3$ star-forming galaxies may show more irregular and clumpy structure compared to local spiral galaxies \citep{Elmegreen2004,Elmegreen2007,Dekel2009a}, further complicating the classification of mergers and isolated disks.

On the other hand, the kinematic properties of galaxies traced by ionized gas or molecular gas have the potential to unambiguously distinguish mergers from isolated disks \citep[e.g.][]{Swinbank2006,Tacconi2006,Menendez-Delmestre2013}.
Integral field spectrograph (IFS) surveys with high spatial/spectral resolutions demonstrate that spiral galaxies often exhibit smooth velocity gradients \citep{Daigle2006,Dicaire2008}, whereas mergers tend to show more complicated kinematic features \citep{Mihos1998,Colina2005}.
Various visual and automatic analysis have also been developed to classify the merger/disk nature of $z\sim1-3$ star-forming galaxies \citep[e.g.][]{Flores2006,Shapiro2008,Hammer2009,Bellocchi2012,Swinbank2012}.

Large and systematic studies of the kinematic properties of galaxies have been made possible with recent IFS surveys of nearby and distant galaxies (e.g. CALIFA survey at $z\sim0$: \citet{Husemann2013}; IMAGES survey at $z\sim0.6$: \citet{Flores2006}; MASSIV survey at $z\sim1$: \citet{Epinat2009,Epinat2012}, SINS survey at $z\sim2$: \citet{Forster-Schreiber2006,Forster-Schreiber2009}).
The samples of galaxies with resolved kinematic information will be further expanded with the ongoing and future IFS surveys such as MaNGA\footnote{https://www.sdss3.org/future/manga.php} and KMOS$^{\rm 3D}$ \citep{Wuyts2014,Wisnioski2014}.
Recent IFS surveys toward $z\sim1-3$ star-forming galaxies (with typical star formation rates (SFR) of $\sim$ 10-500 \Msun\ yr$^{-1}$)  find that $\sim30-50\%$ of these galaxies show evidence of interactions, yet a significant population of these star-forming galaxies ($\sim40-70\%$) have kinematic properties consistent with a rotating disk \citep[e.g.][]{Shapiro2008,Law2009,Wright2009,Epinat2012}.

A possible explanation of the large fraction of rotating disks at $z\sim1-3$ is that these galaxies are formed through rapid, smooth gas accretion \citep[e.g.][]{Keres2005,Birnboim2007,Dekel2009}.
However, some of these disk galaxies also have comparable physical and kinematic properties with systems where the disk is reformed in the merger remnant \citep{Barnes2002,Springel2005b,Robertson2008,Hopkins2009,Narayanan2009}.
In fact, interacting galaxies may display drastically different kinematic features along the merger sequence \citep{Mihos1998,Bellocchi2013}, and it is unclear what the likelihood is that mergers may be mistakenly classified as disk galaxies or vice versa.
For instance, \citet{Bellocchi2012} demonstrate that post-coalescence mergers can be miss-classified as disks when using the kinematic classification criteria designed by \citet{Shapiro2008}.

In this paper, we further explore the robustness of various kinematic classifications for interacting galaxies using a sample of local (U)LIRGs spanning a wide range of merger stages and isolated disk systems.
To make a realistic comparison with the observations of high$-z$ galaxies, we artificially redshift our local sample as if they are observed at $z=1.5$.
We describe our sample and dataset in Section 2.
In Section 3, we describe our analysis including the redshifitng procedures, the spectral line fitting, and the kinematic classification schemes.
We present our results in Section 4 and discuss their implications in Section 5.
We list our conclusions in Section 6.
Throughout this paper, we adopt a $\Lambda$CDM cosmology with $H_0=70$ km s$^{-1}$ Mpc$^{-1}$, $\Omega_{M}=0.3$ and $\Omega_{\Lambda}=0.7$ \citep{Hinshaw2009}.

\section{Sample and Data}
\subsection{IFS Datacubes}
The IFS data of the 24 non-interacting and interacting systems used in this paper are obtained from the Wide Field Spectrosgraph (WiFeS), Integral Field Unit (IFU), and Great Observatory Allsky LIRGs Survey (GOALS) sample \citep[WIGS:][]{Rich2011,Rich2012,Rich2014}.
The GOALS survey \citep{Armus2009} comprises 203 ultraluminous and luminous infrared galaxies ((U)LIRGs; \LIR $\geq 10^{11}$\Lsun)\footnote{\LIR\ $\equiv L_{8-1000\mu m}$ in the object's rest-frame} at $z<0.088$, which is a complete subset of the {\it IRAS} Revised Bright Galaxy Sample \citep[RBGS:][]{Sanders2003}.
The WIGS is a subset of GOALS galaxies that covers a wide range of merger stages and luminosities with a declination upper limit of $+15^{\circ}$ \citep{Rich2012}.

The observations and data reduction are described in detail in \citet{Rich2010,Rich2011,Rich2012}, and here we provide a brief summary.
The IFS data were taken in 2009 and 2010 using WiFeS \citep{Dopita2007,Dopita2010} on the ANU 2.3-meter telescope at the Siding Spring Observatory, and the typical seeing during the observations was $\sim1.5\arcsec$.
Our analysis is restricted to the red channel, with datacubes taken at a spectral resolution of R=7000 and a wavelength coverage of $\sim5500-7000$ \AA. 
Each WiFeS observation consists of 25 slitlets (1\arcsec $\times$ 38\arcsec). 
The WiFeS detectors have a pixel scale of $0.5\arcsec$ along the slit, and the final datacubes are binned by a factor of 2 so that the final pixel size is $1\arcsec \times 1\arcsec$ ($\sim400$ pc at the median $z=0.02$).
The typical total exposure time for each galaxy is 20$-$30 minutes.
All of the final datacubes in this sample consist of at least two individual observations and in some cases a mosaic of multiple pointings to cover the extended structures.

\begin{table*}
\centering
 \caption{List of WIGS Sources}
 \label{tab:data}
 \begin{tabular}{@{}lccccccl}
 \hline
 \hline
IRAS Name & RA (J2000) & Dec (J2000) & $z$ & log(\LIR/\Lsun) & Image/Filter & Int. Stage  & Other Names \\
(1) & (2) & (3) & (4) & (5) & (6) & (7) & (8)  \\
  \hline
F01053$-$1746   &  01:07:47.18 & $-$17:30:25.3  & 0.020067 & 11.71 & ACS F814W & M3 &IC 1623, Arp 236 \\
%F02072$-$1025   & 02:09:42.67 &  $-$10:11:01.9 & 0.012922 &  10.97 & DSS $R$      & ... &NGC 0839, Arp 318 \\
F06076$-$2139  & 06:09:45.81 &  $-$21:40:23.7   & 0.037446 & 11.65 & ACS F814W & M2 & \\
\ \ 08355$-$4944  & 08:37:01.82 &  $-$49:54:30.2   & 0.025898 & 11.62 & ACS F814W & M3 & \\
F10038$-$3338  &10:06:04.80  &  $-$33:53:15.0  & 0.034100  &11.78  & ACS F814W & M3  &  \\
F10257$-$4339 &10:27:51.27 & $-$43:54:13.8   & 0.009354 &11.64  & ACS F814W & M4   &NGC 3256 \\
F12043$-$3140  & 12:06:51.94   & $-$31:56:58.5 & 0.023203 &11.36  & DSS $R$      & M2 & \\
F12592$+$0436 &  13:01:50.80  & $+$04:20:00.0& 0.037483  &11.68  &ACS F814W & M4 &\\
\ \ 13120$-$5453  & 13:15:06.35 & $-$55:09:22.7  &0.030761&12.32  &ACS F814W & M4  &\\
F13373$+$0105 &  13:39:55.00 & $+$00:50:07.0 & 0.022500 & 11.62  &ACS F814W & M2  & NGC 5257/5258, Arp 240 \\
F15107$+$0724 &  15:13:13.09 & $+$07:13:31.8 & 0.012999 & 11.35 & UH-2.2m $I$   & S& \\
F16164$-$0746 &   16:19:11.79 & $-$07:54:02.8 & 0.027152 & 11.62 & ACS F814W & M4  & \\
F16399$-$0937  &   16:42:40.21 & $-$09:43:14.4 & 0.027012& 11.63 & ACS F814W & M3 & \\
F16443$-$2915  & 16:47:31.06 & $-$29:21:21.6  & 0.020881 & 11.37 & IRAC 3.6 \mum    & S  &  \\
F17138$-$1017  &  17:16:35.79 & $-$10:20:39.4 &0.017335 & 11.49   &ACS F814W  & S  & \\
F17207$-$0014  &  17:23:21.95 & $-$00:17:00.9 &0.042810 & 12.46 & ACS F814W& M4  &  \\
F17222$-$5953  &  17:26:43.34 & $-$59:55:55.3 & 0.020781 & 11.41 & DSS $R$      & S  & \\
%\ \ 17578$-$0400  &  18:00:31.90 & $-$04:00:53.3 & 0.014043& 11.48 & DSS $R$     & ...  & \\
F18093$-$5744  &  18:13:39.63 & $-$57:43:31.3 & 0.017345 & 11.62  &  ACS F814W & M2 & IC 4686/4687/4689\\
F18293$-$3413  & 18:32:41.13 & $-$34:11:27.5 & 0.018176 & 11.88 & ACS F814W& M2    \\
F18341$-$5732  &  18:38:25.70 & $-$57:29:25.6 & 0.015611 & 11.35  & DSS $R$      & S  & IC 4734\\
F19115$-$2124  &  19:14:30.90 & $-$21:19:07.0 & 0.048727 & 11.93 & ACS F814W& M2    \\
F20551$-$4250  &  20:58:26.79 & $-$42:39:00.3 & 0.042996 & 12.06 &ACS F814W& M4    \\
F21330$-$3846  & 21:36:10.83 & $-$38:32:37.9 &0.019060 & 11.14 & DSS $R$      & M2  \\
%F21453$-$3511  &  21:48:19.50 & $-$34:57:04.7 & 0.016151 & 11.42 &ACS F606W & mm & NGC 7130, IC 5135\\
F22467$-$4906  &  22:49:39.87 & $-$48:50:58.1 & 0.043033 & 11.84 &ACS F814W & M4   \\
F23128$-$5919  &  23:15:46.78 & $-$59:03:15.6 & 0.044601 & 12.06 &ACS F814W & M3     \\
\hline

 \end{tabular}\par
\smallskip
\begin{flushleft}
{\rm NOTE:} Columns (1) $-$ (5) list the IRAS name, RA, Dec, $z$, and \LIR\ for the WIGS sources, which are compiled from \citet{Sanders2003} and NASA/IPAC Extragalactic Database (NED). 
Column (6) indicates the images and filters that are used for visual morphological classification, column (7) lists the morphological classes from Larson et al. (2014, in prep.), and column (8) lists other common names of WIGS sources.
\end{flushleft}
        
\end{table*}

\subsection{Morphological Classifications}
The interaction stages of the 24 WIGS sources (non-interacting galaxies and interacting systems) were determined by inspection of their optical morphology (Larson et al. 2014, in prep.).
The visual classification scheme developed by Larson et al. (2014) is based on the morphological features during the progression of an interaction sequence \citep[e.g.][]{Barnes1992}: single disk galaxies (S), galaxy pairs without signs of interactions (M1), galaxy pairs with signs of interactions (M2), merged galaxies with double nuclei (M3), merged galaxies with single nucleus and tidal tails (M4), and merged galaxies with an offset single nucleus but without obvious tidal tails (M5).
Comparing to the visual classification scheme developed by \citet{Veilleux2002} and \citet{Surace1998}, M1 stage includes both {\it First approach} and {\it First contact} stages in \citet{Veilleux2002}.
The {\it Pre-merger} stage in \citet{Veilleux2002} is divided into M2 or M3 depending on whether the disk-structure of galaxies can still be seen.
The revised scheme allows us to probe the galaxy kinematics before and after galaxy disks being destroyed during the interaction.

Eighteen of the WIGS sources have been observed with {\it HST} ACS in the $F435W$ band and $F814W$ band with a resolution of 0\arcsec.05 \citep[][Evans et al. in prep.]{Kim2013}, where the high resolution and sensitive {\it HST} images can resolve the detailed structures and reveal the faint interacting features.
The $F814W$ band images are used for morphological classification due to their less severe dust obscuration than the $F435W$ band images.
In fact, dust obscuration may remain an issue at $F814W$ band, which can bias our classification at M3-M4 stages.
Nine out of 12 sources that are classified as M3 or M4 in our sample have NICMOS $F160W$ band images available from \citet{Haan2011}.
The classification results of these 9 sources do not change when $F160W$ band images are used \citep{Haan2011,Petty2014}.
For the other 6 sources that do not have {\it HST} images available, we use $R$ band images from Digitized Sky Survey (DSS), $I$ band images from the UH-2.2 meter telescope \citep{Ishida2004}, or post calibration {\it Spitzer} IRAC 3.6\mum\ images from NASA/IPAC Infrared Science Archive to determine their morphological classes. 
The plate scales of DSS, UH 2.2-meter, IRAC images used in this paper are  1\arcsec, 0\arcsec.22 (seeing$\sim$0\arcsec.9), and 0\arcsec.09 (FWHM$\sim$1\arcsec.66), respectively.
Based on these images, we have 6, 6, 5, 7 sources classified as S, M2, M3, M4, respectively.
Table~\ref{tab:data} lists basic information of galaxies, optical images used for morphological classifications, and the interaction stages.
We note that three WIGS sources include multiple components that are individually resolved in the IFS observations, and we carry out their kinematic classifications separately.
The kinematic classifications of a total of 28 galaxies are listed in Table~\ref{tab:classification}, and their optical images are shown in Figure~\ref{fig:wigs_p1}.

\section{Analysis}
\subsection{Redshifted IFS Datacubes}
To test if the merger/disk classification based on kinematic properties can robustly identify high$-z$ interacting galaxies, we artificially redshift the IFS datacubes of local galaxies described in Section 2.1.
Current high$-z$ IFS surveys cover a range of redshifts from $z\sim1-3$ \citep{Forster-Schreiber2009,Law2009,Wright2009,Wisnioski2014}.
For example, the SINFONI sample from the MASSIV survey has a median redshift of $z\sim1.2$ \citep{Epinat2009,Epinat2012}.
Recent KMOS$^{\rm 3D}$ survey focuses on star-forming galaxies at $z\sim1$ and $z\sim2$ \citep{Wisnioski2014}, and the AMAZE and LSD projects target galaxies at $z\sim3$ \citep{Gnerucci2011}.
In this paper, we artificially redshift our local galaxies to $z=1.5$, as a case study.
The redshift is chosen to match a lensed spiral galaxy at $z=1.5$ (Sp1149), whose fully reduced Keck/OSIRIS observations are available to simulate the sky-line dominated noise of near infrared IFS observations \citep{Yuan2011}.
The corresponding physical sizes of a given angular scale only differ by $\sim1\%$ between $z=1.5$ and $z=2$. 
Thus our choice of $z=1.5$ should be representative of typical $z\sim1-2$ observations in terms of the spatial resolutions can be achieved.
Detail investigations of the redshift dependence regarding to the sky line distributions and instrument configurations are beyond the scope of this paper.
We assume the observations are carried out under seeing-limited condition ($0\arcsec.5$) with pixel size of $0\arcsec.1\times 0\arcsec.1$ ($\sim 900$ pc at $z=1.5$).

To redshift the local galaxies, we first define an elliptical mask for each galaxy.
This is to prevent contamination from foreground stars and other artifacts when binning the pixels and convolving with larger beams.
We then follow the redshifting procedure of IRASF17222$-$5953 described in \citet[][Section 4.2]{Yuan2013}.
We derive the surface brightness distribution according to the redshift and plate scales of the input (WiFeS) and output (OSIRIS) datacubes.
For instance, the WiFeS plate scale (1\arcsec$\times$1\arcsec) corresponds to $\sim$400 pc $\times$ 400 pc at $z=0.02$, and we then bin the pixels by a factor of two to match the corresponding physical sizes of the $0\arcsec.1\times 0\arcsec.1$ plate scale at $z=1.5$ ($\sim$900 pc $\times$ 900 pc).

Then we use the source-free region of the Sp1149 data to create noise datacubes with the dimensions of the spatially binned and dimmed WiFeS data.
The noise datacube is interpolated to match the spectral resolution of WiFeS data.
We combine the spatially-binned WiFeS data with the noise datacube, and then convolve the resulting datacube with a two-dimensional Gaussian (FWHM=0\arcsec.5), a resolution comparable to the typical seeing-limited observations of $z\sim1-3$ star-forming galaxies.   
Since the emission lines of these degraded data become too faint to be detectable, we apply an artificial brightening factor to ensure that the peak of \Ha\ emission has a signal-to-noise (S/N) ratio of 50.

\subsection{Spectral Line Fitting}
To extract the kinematic information from the artificially redshifted datacube, we fit Gaussian profiles to \Ha\ ($\lambda_{rest-frame} = 6562.8$ \AA\ and two adjacent \NII\ lines ($\lambda_{rest-frame}=6548,6583$ \AA\ simultaneously.
We fix the flux ratio between \NII$\lambda6583$ and \NII$\lambda6548$ at 3 \citep{Osterbrock1989}, and we assume the linewidths of the \Ha\ and \NII\ lines are the same.
An instrumental profile of $0.94$ \AA\ is then subtracted from the measured linewidth and the final linewidth $=(\sigma_{measured}^2-\sigma_{instrument}^2)^{0.5}$.
In the end, our fitting profile contains five free parameters: systematic velocity, spectral linewidth, \Ha\ flux, \NII$\lambda6548$ flux, and the continuum level ($\lambda6400-6750$ \AA).

We use the {\tt IDL} routine {\tt MPFIT} to perform the spectral line fitting \citep{Markwardt2009}, which uses the Levenberg-Marquardt algorithm to solve non-linear least-squares problems. 
We repeat the spectral line fitting at each pixel across the entire map, and each line fitting is weighted based on the inverse squared noise spectrum.
The uncertainties of each free parameter are the 1 $\sigma$ errors computed from the covariance matrix in the {\tt MPFIT} routine.
We use the \Ha\ line as a tracer of a galaxy's kinematics.
A minimum of S/N=5 is required for the \Ha\ line to be considered a robust detection \citep[e.g.][]{Yuan2013}.
No significant differences are seen in the bins when using adaptive spatial binning \citep{Cappellari2003}.
Based on the best fit systematic velocity and line width at each pixel, we then construct the velocity distribution map and the velocity dispersion map.
An example galaxy (IRASF01053$-$1746) is shown in Figure~\ref{fig:wigs_single}, and we present the entire sample in the Appendix (Figure~\ref{fig:wigs_p1}).

\begin{figure*}
\centering
  \includegraphics[width=0.85\textwidth]{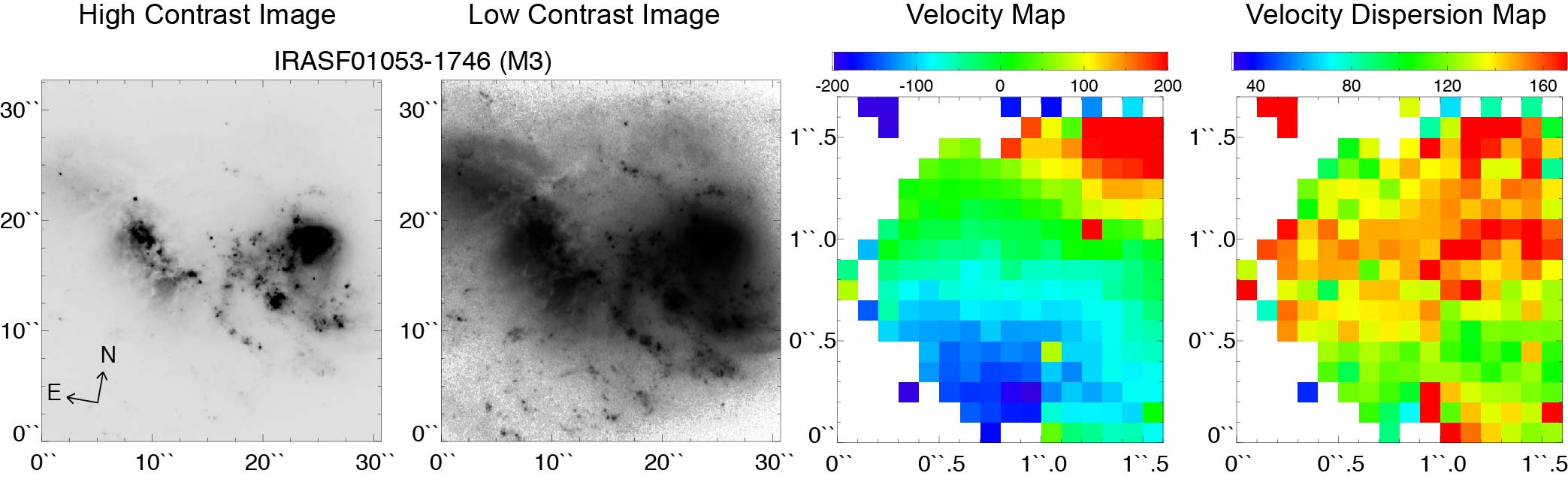} 
\caption{The left two panels shows the ACS/$F814W$-band image of IRASF01053$-$1746 with two different stretching schemes (inverse hyperbolic sine and histogram equalization) and its morphological classification (M3).
The right two panels show the velocity map and velocity dispersion map derived from the redshifted IFS datacubes.
The velocity and velocity dispersion maps have the same field-of-view in physical scale as the optical images shown in the left panel, and their angular sizes are scaled to $z=1.5$.
The color bars are in the unit of km s$^{-1}$.
We present the maps for the entire sample in the Appendix (Figure~\ref{fig:wigs_p1}). 
} 
\label{fig:wigs_single}
\end{figure*}

\begin{table*}
\centering
 \caption{Results of Kinematic Classifications}
 \label{tab:classification}
 \begin{tabular}{@{}lcccc}
 \hline
 \hline
                      &                  & {\it Kinemetry} in S08 & Revised {\it Kinemetry} in B12  & Visual Kinematic\\
IRAS Name     & Int. Stage    &$K_{asym}$ (Classification) &   $K_{asym,B12}$ (Classification) & Classification  \\
  \hline
F01053$-$1746  & M3 & 0.16$\pm$0.04 (Disk) &  0.20$\pm$0.05 (Merger) & RD\\
F06076$-$2139  & M2 & 1.82$\pm$0.52 (Merger) & 1.38$\pm$0.43 (Merger) & CK   \\
\ \ 08355$-$4944 & M3 &4.06$\pm$1.80 (Merger) & 4.39$\pm$1.81 (Merger)  & CK \\
F10038$-$3338  & M3 & 5.18$\pm$2.39 (Merger) & 7.23$\pm$2.92 (Merger)   &  CK \\
F10257$-$4339  & M4 & 0.05$\pm$0.01 (Disk) & 0.04$\pm$0.03 (Disk)  &  RD \\
F12043$-$3140  & M2 &  0.27$\pm$0.07 (Disk)  & 0.34$\pm$0.07 (Merger)  & PD  \\
F12592$+$0436  & M4 & 0.15$\pm$0.01 (Disk)  & 0.12$\pm$0.01 (Disk)  &   RD \\
\ \ 13120$-$5453  & M4 & 0.39$\pm$0.08 (Disk) & 0.33$\pm$0.07 (Merger)  &  RD\\
F13373$+$0105 W & M2 & 0.09$\pm$0.01 (Disk) &  0.07$\pm$0.01 (Disk)  & RD  \\
F13373$+$0105 E & M2 & 0.14$\pm$0.02 (Disk) & 0.12$\pm$0.01 (Disk)  &  RD  \\
F15107$+$0724  & S & 0.50$\pm$0.06 (Disk)  & 0.52$\pm$0.05 (Merger)   &  PD \\
F16164$-$0746  & M4 & 2.31$\pm$0.77 (Merger) & 2.38$\pm$0.77 (Merger)  & CK \\
F16399$-$0937  & M3 & 0.21$\pm$0.06 (Disk) & 0.20$\pm$0.06 (Merger)  &  PD\\
F16443$-$2915 S & S &0.34$\pm$0.06 (Disk) & 0.41$\pm$0.07 (Merger)  & PD \\
F16443$-$2915 N & S & 0.10$\pm$0.01 (Disk) & 0.07$\pm$0.01 (Disk)  &  RD \\
F17138$-$1017  & S & 0.29$\pm$0.04 (Disk) & 0.27$\pm$0.03 (Merger)  & PD \\
F17207$-$0014  & M4 & 0.19$\pm$0.08 (Disk) & 0.26$\pm$0.10 (Merger)  & PD \\
F17222$-$5953  & S & 0.11$\pm$0.02 (Disk)  & 0.13$\pm$0.02 (Disk)  &  RD \\
F18093$-$5744 N  & M2 & 0.11$\pm$0.02 (Disk) &  0.11$\pm$0.01 (Disk)  &  RD  \\
F18093$-$5744 S  & M2 & 0.18$\pm$0.03  (Disk) & 0.21$\pm$0.04 (Merger)  & RD   \\
F18093$-$5744 C  & M2 & 0.76$\pm$0.10 (Merger) & 0.55$\pm$0.12 (Merger)  & CK  \\
F18293$-$3413  & M2 & 0.24$\pm$0.06 (Disk) & 0.26$\pm$0.07 (Merger) &  PD \\
F18341$-$5732  & S & 0.10$\pm$0.01 (Disk) & 0.09$\pm$0.01 (Disk) &  RD  \\
F19115$-$2124  & M2 & 0.57$\pm$0.24 (Merger) & 0.93$\pm$ 0.32 (Merger) & CK   \\
F20551$-$4250  & M4 & 0.31$\pm$0.03 (Disk) & 0.25$\pm$0.02 (Merger)  &  PD \\
F21330$-$3846  & M2 & 0.41$\pm$0.24 (Disk)  & 0.60$\pm$0.31 (Merger)  &  PD  \\
F22467$-$4906  & M4 & 0.82$\pm$0.36 (Merger) & 0.62$\pm$0.27 (Merger)  &  CK \\
F23128$-$5919  & M3 & 0.43$\pm$0.06 (Disk) &  0.48$\pm$0.07 (Merger) &  CK \\
\hline
     
 \end{tabular}\par
\smallskip
\begin{flushleft}
{\rm NOTE:} Kinematic classifications of S, M2, M3, and M4 galaxies based on the classification schemes described in Section 3.3.1-3. The merger/disk classification criteria in the \citet{Shapiro2008} scheme follows $K_{asym}>0.5$, and the classification criteria in the \citet{Bellocchi2012} follows $K_{asym, B12}>0.146$. 
\end{flushleft}

\end{table*}

\subsection{Kinematic Classifications}

Several kinematic classification schemes have been developed to classify merger/non-merger of local and $z\sim1-3$ star-forming galaxies \citep[e.g.][]{Flores2006,Shapiro2008,Goncalves2010,Alaghband-Zadeh2012,Bellocchi2012,Epinat2012,Bellocchi2013}.
Here we use the classification schemes based on kinematic asymmetries developed by \citet[][hereafter S08]{Shapiro2008} and revised by \citet[][hereafterB12]{Bellocchi2012}, as well as a modified version of the visual classification scheme developed by \citet{Flores2006}.

\subsubsection{Kinematic Asymmetries in \citet{Shapiro2008}}

The kinemetry analysis \citep{Krajnovic2006} is designed for modeling the higher-order moments of the velocity and velocity dispersion distributions of galaxies.
The line-of-sight velocity map or velocity dispersion map $K(a,\psi)$ can be divided into several elliptical rings (with semi-major axis $a$) as velocity or velocity dispersion profiles.
These profiles can then be described as an expansion of $N+1$ harmonic terms:
\begin{equation}
K(a,\psi)=A_0(a)+\sum\limits_{n=1}^N A_n(a) \sin{n\psi}+B_n(a) \cos{n\psi},
\end{equation}
where $\psi$ is the azimuthal angle.
For an ideal disk, its velocity field and velocity dispersion field are dominated by $B_{1}$ and $A_{0}$, and thus higher-order deviations may be a result of galaxy mergers.
S08 quantify the level of deviation from an ideal disk by defining asymmetric measures of velocity and velocity dispersion fields ($v_{asym}$ and $\sigma_{asym}$) using the higher order kinematic coefficients (Equation 4 and 5 in S08).
%\begin{equation}
%v_{asym} = \left< \frac{\sum\limits_{n=2}^5 k_{n,v}/4}{B_{1,v}} \right>_r,
%\sigma_{asym} = \left< \frac{\sum\limits_{n=1}^5 k_{n,\sigma}/5}{B_{1,v}} \right>_r,
%\end{equation}
%where $k_n=(A_n^2+B_n^2)^{1/2}$, the subscripts $v$ and $\sigma$ refer to the quantifies corresponding to the velocity and velocity dispersion maps, and $r$ refers to the average over all radius steps.
Based on the templates from local galaxies and simulated galaxies, S08 define a criterion to identify mergers via $K_{asym}=(v_{asym}^2+\sigma_{asym}^2)^{1/2}>0.5$.

We measure $v_{asym}$ and $\sigma_{asym}$ of our redshifted WIGS galaxies using the {\tt IDL} routine {\tt Kinemetry} \citep{Krajnovic2006}.
Before performing the kinemetry analysis, we determine the position of the galaxy center based on the distribution of the continuum emission.
The continuum map is first smoothed using a Gaussian with FWHM of 3 pixels (corresponds to 0\arcsec.3 in the redshifted datacubes), and then the galaxy center is determined as the centroid of the smoothed map (typical difference of using different smoothing factors of 1-5 pixels is less than 1 pixel).
We note that the smoothed continuum maps are only used for determining galaxy center.
We avoid using the \Ha\ emission since it traces the recent star forming regions and may not necessarily represent the overall stellar distribution.
As discussed in S08 and B12, the derivation of kinematic asymmetries is sensitive to the position of the galaxy center determination.
We discuss how the accuracy of the galaxy center impacts our results and conclusions in Section 4.3.

After fixing the center, {\tt kinemetry} then finds the best-fit ellipse with position angle (PA) and the flattening factor ($Q=1-e$) at each radius step until more than 25\% of the data points along an ellipse are not present.
S08 derive a global PA and $Q$ before performing the kinemetry analysis, but here we treat them as free parameters since the meaning of a global PA and $Q$ for mergers is unclear.
B12 also show that the choice of either free or fixed PA or $Q$ does not have a strong impact on their results.
We derive $K_{asym}$ of six single disk galaxies by attempting to find a global PA and Q prior to the kinemetry analysis, none but one galaxy (F16443$-$2915 S) alters the classification result.

\subsubsection{Revised Kinematic Asymmetries in \citet{Bellocchi2012}}

Although the local merger templates used in S08 include 3 interacting pairs, 2 double nuclei mergers, and 3 single nucleus mergers, the sample size remains small and thus they may not trace the full scheme of the interaction sequence.
B12 find that their sample of both non-interacting disks (IRASF10567$-$4310 and IRASF11255$-$4120) and post-coalescence mergers (IRASF04315$-$0840 and IRASF21453$-$3511) are classified as disks using the kinematic classification scheme developed by S08.
Since a post-coalescence merger may be dominated by rotation in the inner regions but retain distorted kinematics at the outskirts \citep[e.g.][]{Kronberger2007}, B12 modify the asymmetric measures defined in S08 by weighting the asymmetries according to the circumferences of each radius step (Equation (6) and (7) in B12; hereafter indicated as $v_{asym,B12}$ and $\sigma_{asym,B12}$).
%\begin{equation}
%v_{asym,B12} = \frac{\sum\limits_{r=1}^N\left(\frac{k_{ave,v}}{B_{1,v}} \times C_{r,v}\right)}{\sum\limits_{r=1}^N C_{r,v}},
%\sigma_{asym,B12} = \frac{\sum\limits_{r=1}^N\left(\frac{k_{ave,\sigma}}{B_{1,v}} \times C_{r,\sigma}\right)}{\sum\limits_{r=1}^N C_{r,\sigma}},
%\end{equation}
%where $k_{ave,v}=\sum\limits_{n=2}^5 k_{n,v}/4$ and $k_{ave,\sigma}=\sum\limits_{n=1}^5 k_{n,\sigma}/5$.
%The circumferences $C$ of each radius step is
%\begin{equation}
%C(e,a) \approx 2\pi a \left[ 1-\left( \frac{1}{2}\right) ^2 e^2-\left(\frac{3}{8}\right)^2\times\frac{e^4}{3}\right] ,
%\end{equation}
%where $a$ is the semi-major axis of ellipse and $e$ is the ellipticity.
Based on the revised measurements of kinematic asymmetries of two non-interacting disks and two post-coalescence mergers, B12 derive a new criterion to classify mergers via $K_{asym,B12}=(v_{asym,B12}^2+\sigma_{asym,B12}^2)^{1/2}>0.146$.

\subsubsection{Visual Kinematic Classification}

In addition to the two automatic classification schemes defined by S08 and B12, we also visually classify the dynamical status of galaxies based on the distribution of their velocity and velocity dispersion fields.
We adopt the 3-class kinematic classification scheme developed by \citet[][also see \citet{Bellocchi2013}]{Flores2006} with some modifications.
In \citet{Flores2006} and \citet{Bellocchi2013}, comparison between kinematic maps and optical morphology (alignments between optical major axis and velocity gradient) is required to determine the kinematic classes of galaxies.
To make a direct comparison with the kinemetry-based classifications, which do not require the alignment with optical major axis, we modify the visual classification schemes in \citet{Flores2006} and \citet{Bellocchi2013} by removing the criteria that use optical morphology.

Below we list the modified classification criteria of three kinematic classes determined based on a qualitative assessment of galaxies' velocity and velocity dispersion maps (Figure~\ref{fig:wigs_p1}):
\begin{enumerate}
\item Rotating disk (RD): the velocity map shows a clear velocity gradient, and the velocity dispersion map should show a clear peak near the galaxy center where the gradient of the rotation curve is the steepest \citep[e.g.][]{van-Zee1999}.

\item Perturbed disk (PD): the velocity map shows a clear velocity gradient as RD whereas the velocity dispersion map shows a peak offset from the center or does not show a clear peak. 

\item Complex kinematics (CK): both velocity and velocity dispersion maps do not follow the expected pattern of normal rotating disks.

\end{enumerate}

\begin{figure*}
\centering
  \includegraphics[width=0.9\textwidth]{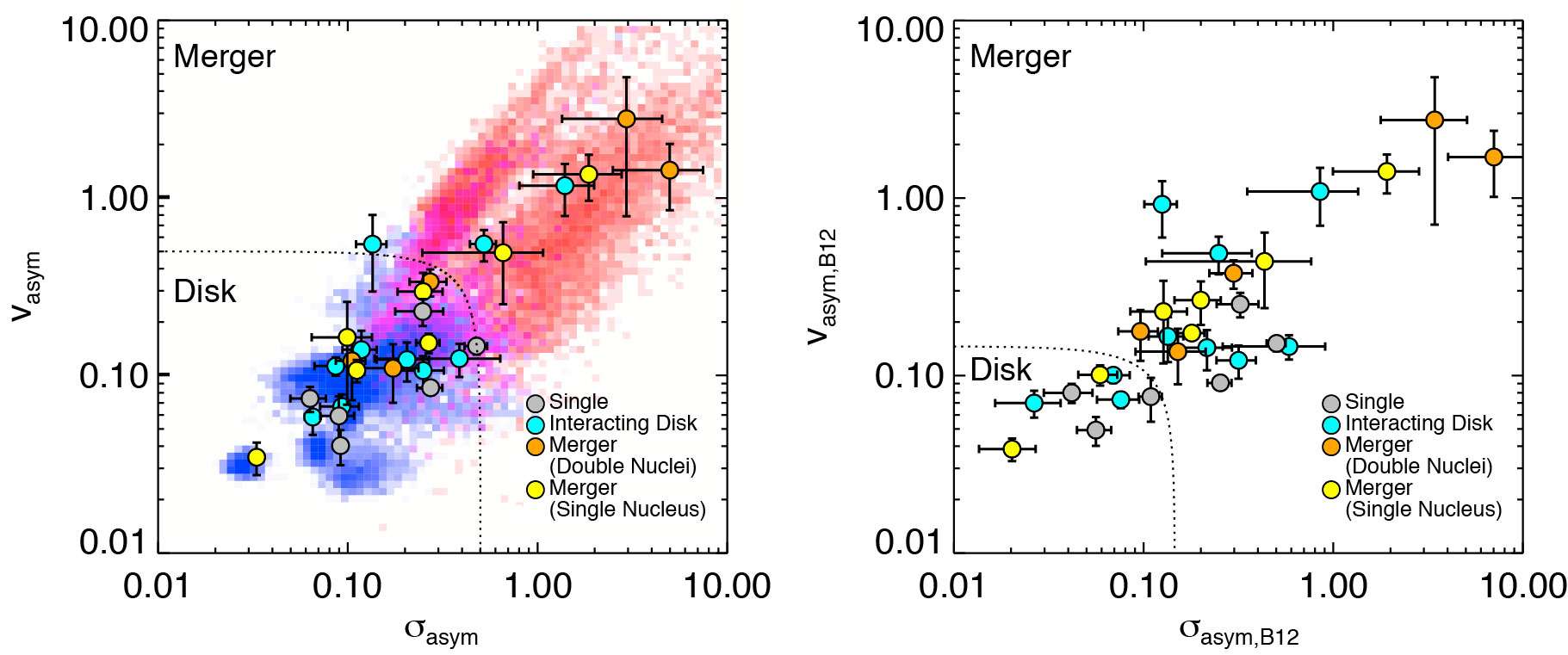} 
\caption{The distribution of kinematic asymmetries for our redshifted WIGS galaxies (with a spatial resolution of 0\arcsec.5). 
The left panel shows the results based on $v_{asym}$ and $\sigma_{asym}$ defined by S08.
Data points are overlaid on the merger (red) and disk (blue) templates distribution in S08, and the black dotted line refers to the merger/disk classification criteria of $K_{asym}=(v_{asym}^2+\sigma_{asym}^2)^{1/2}=0.5$. 
The right panel shows the results based on the revised asymmetries defined by B12, and the black dotted line refers to the merger/disk classification criteria of $K_{asym,B12}=(v_{asym,B12}^2+\sigma_{asym,B12}^2)^{1/2}=0.146$.
In both panels, data points are color-coded according to their morphological classifications, in which the classification scheme is defined based on the progression of an interaction sequence: gray (S: single disk galaxies), cyan (M2: interacting disks), orange (M3: mergers with double nuclei and tidal tails), and yellow (M4: mergers with single nucleus and tidal tails). 
The error bars represent the standard deviation of the mean $v_{asym}$ and $\sigma_{asym}$ across all radius steps.
} 
\label{fig:kinemetry}
\end{figure*}

\section{Results}

\subsection{Kinemetry-based classifications along the interaction sequence}

All the kinematic classifications of S, M2, M3, and M4 galaxies based on the classification schemes in Section 3.3 are listed in Table~\ref{tab:classification}.
Figure~\ref{fig:kinemetry} shows the distribution of kinematic asymmetries defined by S08 and B12 for our WIGS galaxies, which are artificially redshifted to $z=1.5$ assuming seeing-limited conditions (FWHM=0\arcsec.5).
The data points are color-coded according to their interaction stages and the error bars represent the standard deviation of the mean $v_{asym}$ and $\sigma_{asym}$ across all radius steps.
When the S08 scheme is used, all 6 single disk galaxies are classified as disks based on the kinematic asymmetries, yet only 8 out of 22 interacting systems are classified as mergers. 
As an example, IFASF12592+0436 demonstrates the case that the morphological and kinematic classifications are inconsistent.
Despite the obvious tidal tails in the $F814W$-band image, the velocity and velocity dispersion fields of IFASF12592+0436 do not show complicated structures as one might expect for a merger.

\begin{figure}
  \includegraphics[width=0.5\textwidth]{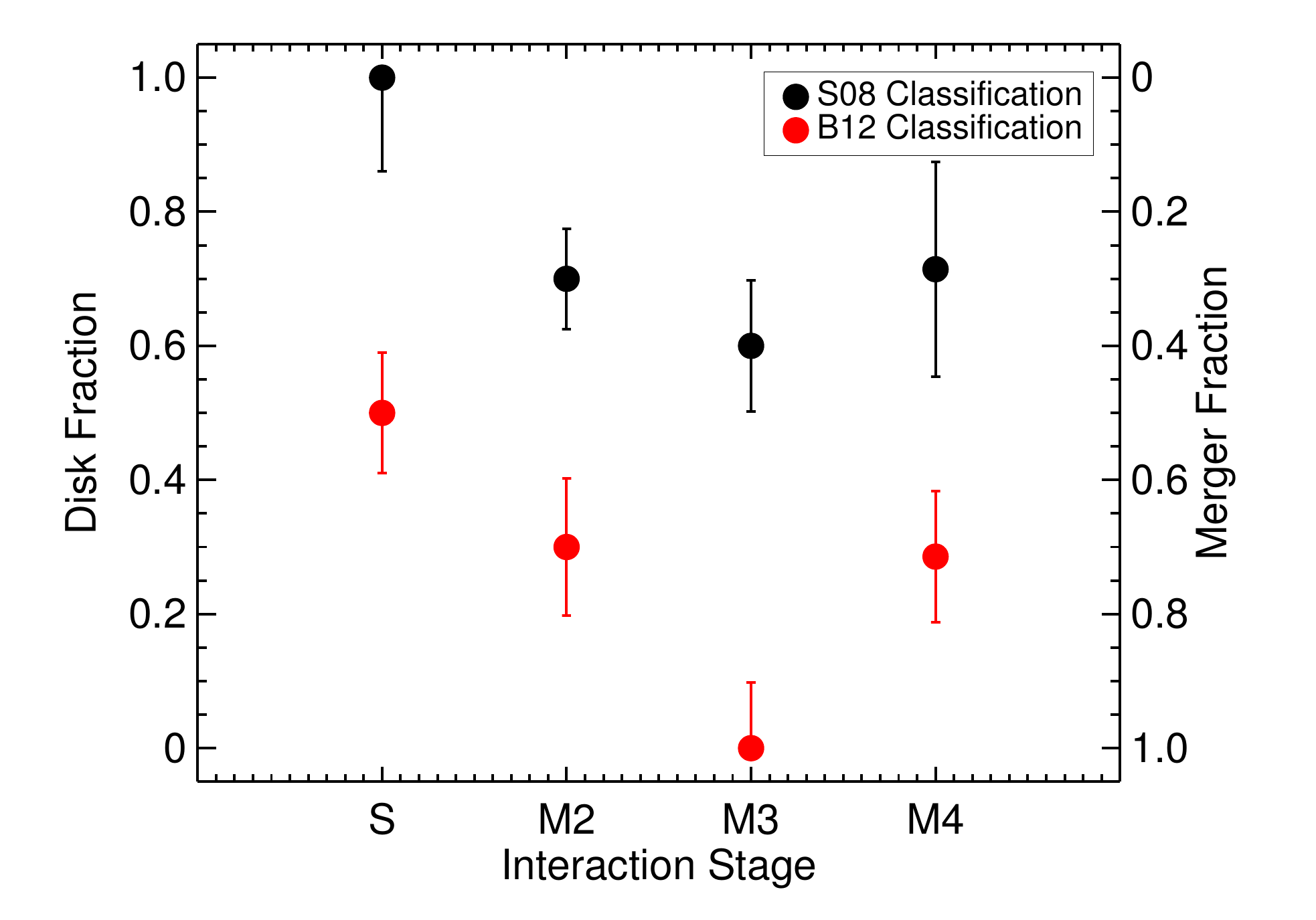} 
\caption{The fraction of galaxies to be classified as disks/mergers based on the measurement of kinematic asymmetries as a function of interaction stages (S: single disk galaxies, M2: interacting disks, M3: mergers with double nuclei, and M4: mergers with single nucleus).
The black dots are the results based on the criteria defined by S08 and the red dots are based on the criteria defined by B12.
Error bars are derived based on the frequency (out of four additional testing cases) of altering the merger/disk classification when shifting the center of galaxies horizontally and vertically by 1 pixel.
} 
\label{fig:class_s08b12}
\end{figure}

The consistency between merger/disk classifications based on morphological features and kinematic asymmetries shows a strong trend with galaxy interaction stage.
Figure~\ref{fig:class_s08b12} shows the fraction of galaxies that are classified as disks and mergers based on kinematic asymmetries as a function of morphological classifications.
When the classification criteria of S08 is used, 100\% of single disk galaxies in our sample are classified as disks, whereas only 30-40\% of the interacting systems are classified as mergers.
Since the interacting disks (M2) still show clear disk structures in spite of their signs of interactions, these galaxies may retain disk-like kinematics. 
However, our sample of later-stage mergers (mergers with double nuclei and single nucleus; M3 and M4) display highly disturbed morphological features, yet only $\sim30-40$\% are classified as mergers.

On the other hand, when the classification scheme based on the revised kinematic asymmetries in B12 is used, all of our M3 galaxies are classified as mergers, whereas the success rate for single disk galaxies to be classified as disks decreases to only $\sim50\%$.
The differences in the definition of kinematic asymmetries (by simply averaging across all radius steps or weighted according to the circumferences) may contribute to the differences of merger/disk fractions measured using the S08 or B12 scheme.
However, the median difference of the $K_{asym}$ and $K_{asym,B12}$ values is only $\sim16\%$ for our galaxies (Table~\ref{tab:classification}), whereas the classification criteria used in B12 ($K_{asym,B12}$=0.146) is more than three times lower compared to S08 ($K_{asym}$=0.5).
Thus the significantly lower classification threshold used in B12 is the main reason of the drastically different disk and merger fractions derived using the S08 and B12 schemes.
We note that B12 shows that the classification criterion between mergers and disks decreases from 0.146 to 0.13 when artificially redshifting their datacubes to $z=3$, and thus a lower threshold than 0.146 should be applied to our redshifted datacubes.
However, none of our classification results changes using a criterion of either 0.146 or 0.13.
Figure~\ref{fig:kinemetry} demonstrates that there is a significant overlap between interacting systems and single galaxies in the distribution of kinematic asymmetries using either S08 or B12 scheme, which implies that a single $K_{asym}$ or $K_{asym,B12}$ cut is unable to differentiate interacting systems from single galaxies.
A higher threshold can lead to an underestimation of merger fractions whereas a lower threshold can overestimate merger fractions.

The results on the reliability of merger/disk classifications based on kinematic properties rely on robust morphological classifications. 
Six galaxies (4 S galaxies, 2 M2 galaxies) included for kinematic classifications have only low resolution DSS or IRAC images available (with the resolution $>$1\arcsec), and their morphological classifications are less reliable compared to the other 22 galaxies.
If we exclude the six galaxies with low resolution images, only 2 S galaxies remain in our sample.
Both of these two S galaxies are classified as disks when the S08 scheme is used, and are classified mergers when the B12 scheme is sued.
For M2 galaxies, the merger fraction increases from 30\% to 37.5\% when the S08 scheme is used, and decreases from 70\% to 62.5\% when the B12 scheme is used.

\subsection{Comparison between visual kinematics and kinemetry-based classifications}

\begin{figure}
  \includegraphics[width=0.5\textwidth]{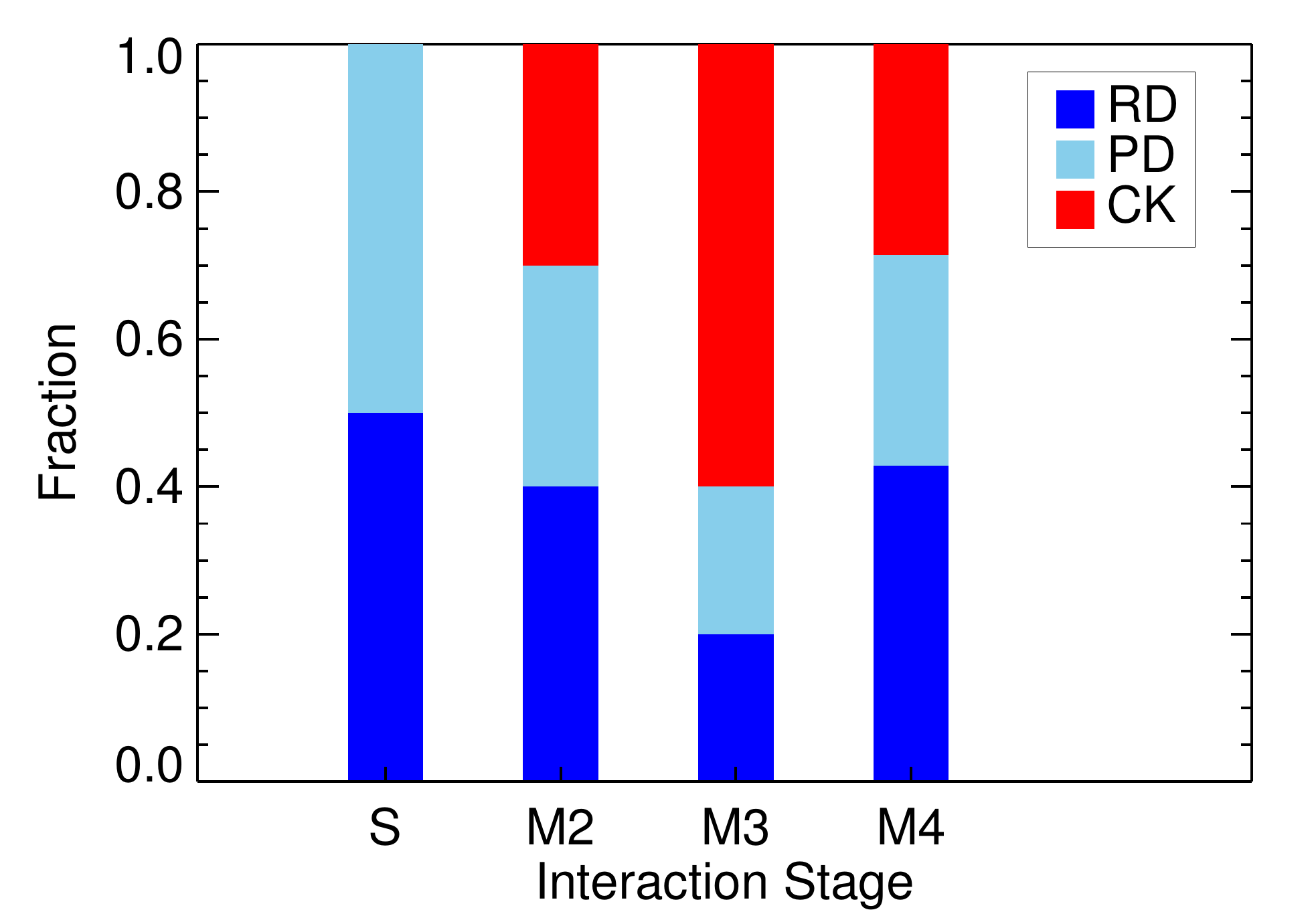} 
\caption{The fraction of S, M2, M3, and M4 galaxies to be classified as rotating disks (RD), perturbed disks (PD), and complex kinematics (CK).
} 
\label{fig:class_f06}
\end{figure}

Figure~\ref{fig:class_f06} shows the classification results based on the visual kinematic classification described in Section 3.3.3.
The distribution of rotating disks (RD), perturbed disks (PD), and complex kinematics (CK) along the interaction stages is consistent with the trend based on kinemetry-based classifications.
All of our single disk galaxies are classified as either RD or PD.
As the interaction stage progresses, higher fraction of the galaxies begin to show complicated kinematic features.
For instance, 60\% of the M3 galaxies are classified as CK.
However, a significant population of M2, M3, and M4 galaxies retain an ordered velocity gradient in their velocity maps and thus are classified as either RD or PD.
We compare our visual kinematic classification results with  \citet{Bellocchi2013} based on the overlapped sample of 11 galaxies.
As noted in Section 3.3.3, the visual classification scheme used in \citet{Bellocchi2013} requires the comparison with optical major axis.
In spite of this difference and that our redshifted datacubes have a factor of 2-6 worse spatial resolution in physical scales compared to \citet{Bellocchi2013}, seven out of these 11 galaxies have consistent visual kinematic classification results, and the discrepancies in the other 4 galaxies are mainly due to the degree or complexity (PD or CK) or the requirement of alignment with optical major axis (RD or PD). 

Table~\ref{tab:comparison} compares the results between visual kinematics and kinemetry-based classifications.
When S08 criteria is used, all galaxies identified as RD are classified as disks and all but one galaxies identified as CK are classified as mergers.
When B12 criteria is used, all but three identified as RD are classified as disks and all galaxies identified as CK are classified as mergers.
The major difference is those galaxies that are identified as PD, where all of them are classified as disks in S08 criteria but classified as mergers in B12 criteria.
The comparison results show that different kinematic classification schemes can lead to large dispersions in the population of identified mergers.

\begin{table}
\vspace{8pt}
\centering
 \caption{Comparison between visual kinematics and kinemetry-based classifications}
 \label{tab:comparison}
 \begin{tabular}{@{}lccc}
 \hline
 \hline
 Visual Classes& RD & PD & CK \\
 \hline
 S08 Disks   & 11 & 9 & 1  \\
 S08 Mergers & 0 & 0 & 7 \\
 \hline
 B12 Disks   & 8 & 0  & 0 \\
 B12 Mergers & 3 & 9  & 8 \\
 \hline
 
 \end{tabular}\par
\end{table}

%\begin{table*}
%\vspace{8pt}
%\centering
% \caption{Comparison between visual and {\it kinemetry}-based classifications}
% \label{tab:comparison}
% \begin{tabular}{@{}lccccccccccccccc}
% \hline
% \hline
% Interaction Stages    &    & S  &   & &   & M2  &   &   & & M3  &   & &   & M4  &   \\
% Visual Classes        & RD & PD & CK& &RD & PD & CK& &RD & PD & CK& &RD & PD & CK \\
% \hline
% S08 Disks   & 3 & 3 & 0 & & 4 & 3 & 0 & & 1 & 1 & 1 & & 3 & 2 & 0\\
% S08 Mergers & 0 & 0 & 0 & & 0 & 0 & 3 & & 0 & 0 & 2 & & 0 & 0 & 2 \\
% \hline
% B12 Disks   & 3 & 0  & 0 & & 3 & 0 & 0 & & 0 & 0 & 0 & & 2 & 0 & 0\\
% B12 Mergers & 0 & 3  & 0 & & 1 & 3 & 3 & & 1 & 1 & 3 & & 1 & 2 & 2\\
% \hline
% \end{tabular}\par
%\end{table*}

\subsection{Uncertainties in the kinemetry-based classifications}

As mentioned in Section 3.3.1, the derivation of $v_{asym}$ and $\sigma_{asym}$ is sensitive to the position of the galaxy center.
The uncertainties of galaxy center can be a significant factor when different determination methods are used.
For instance, whether we use the position of the peak emission or derive the centroid with different smoothing factors (see the description in Section 3.3.1).
We derive the difference between the position of the center determined in Section 3.3.1 and the cases that are determined based on the peak of continuum maps or another four different smoothing factors (1,2,4,5 pixels), and we find that the average difference is 0.45 pixel (with the standard deviation of 0.44 pixel).  
In fact, one of our WIGS sample overlaps with the merger templates used in S08 (IRASF17207$-$0014), yet our derived $K_{asym}$ is $\sim6$ times smaller compared to the median $K_{asym}$ of the Monte Carlo realizations in S08.
Although the differences between the datacubes used in S08 and in this paper may explain the difference in $K_{asym}$, the most likely explanation is the difference in galaxy center (the center determined by S08 is located close to the edge of the IFU datacube; Figure 9 in S08).
A similar value of $K_{asym}$ as S08 is derived when shifting the galaxy center by 1-2 pixels.

Since the pixels are discretized (each pixel corresponds to 0\arcsec.1 and $\sim$0.9 kpc in the redshifted datacubes), we estimate the significance of galaxy center determination to our conclusions by calculating the kinematic asymmetries in four additional cases for all galaxies assuming the galaxy center is shifted by one pixel horizontally or vertically with respect to the center determined based on the continuum map.
We determine the uncertainties in fraction of galaxy classification results (Figure~\ref{fig:class_s08b12}) based on the likelihood that the merger/disk classification alters when shifting the galaxy center.
The resulting uncertainties are typically $10-20\%$, and thus our conclusions remain after considering the uncertainty in the determination of galaxy center.

\section{Discussion}

By carrying out kinematic classifications of a set of 24 interacting systems and non-interacting galaxies, we find that mergers with two separate nuclei (M3 galaxies) are the most easily distinguished from single disk galaxies, and a higher fraction of early and late stage mergers (M2 and M4) have kinematic properties that can be confused with rotating disks.

The high fraction of interacting galaxies showing kinematics consistent with disks is in disagreement with the conclusions based on local merger templates in S08, and this differences cannot be explained by the uncertainties in the galaxy center determination alone.
However, one local merger template (IRAS12112$+$0305, out of eight galaxies) in S08 that shows disk-like kinematics has been excluded for deriving the merger/disk classification criteria.
If IRAS12112$+$0305 was included by S08 in determining merger/disk $K_{asym}$ thresholds, the differences of the derived disk fractions between this work and S08 may be smaller.
Furthermore, both of the WIGS and S08 samples are of modest sizes, and these samples may cover a different range of interaction sequence.
For instance, we also include interacting pairs that are individually resolved in the IFS observations, whereas the three interacting pairs used S08 are all unresolved.

As discussed in Section 4.1, our M2 galaxies show clear disk structure in the optical images, and thus we expect them to retain disk-like kinematics.
Five out of six M2 systems that are individually resolved in the IFS observations F13373$+$0105 W/E, F18093$-$5744 N/S, F18293$-$3413, exception: F18093$-$5744 C) show significantly lower kinematic asymmetries calculated using either S08 or B12 schemes compared to the other four M2 systems that are unresolved in the IFS observations (F06076$-$2139, F12043$-$3140, F19115$-$2124, F21330$-$3846).
This difference suggests that individually resolved galaxies in the interacting pairs tend to show disk-like kinematics, whereas if the two galaxies are unresolved, then they tend to show more complicated kinematics.
For example, the unresolved dynamical information from two galaxies in F19115$-$2124 leads to its highly disturbed and complicated velocity field.

One possible origin of the disk-like kinematics in the late-stage mergers (M4) is the disks reformed in gas-rich merger remnants \citep[e.g.][]{Barnes2002,Springel2005b,Robertson2008}.
Some of the local (U)LIRGs may have comparable gas fractions to those of the gas-rich merger simulations \citep[e.g.][]{Sanders1991} that are necessary to reform the gas disk in galaxy mergers \citep[e.g.][]{Downes1998,Tacconi1999}.
The emergence of these gas disks and their subsequent formation of stellar disks are also observed in luminous mergers and  merger remnants \citep[e.g.][]{Rothberg2010,Medling2014}.
Recently, \citet{Ueda2014} study the $^{12}$CO kinematic properties of optically-selected local merger remnants from \citet{Rothberg2004} (morphological properties consistent with our M4 and M5 stages), and they find that 80\% (24/30) of their sample show kinematic signatures of rotating molecular gas disks.
Such high disk fraction is consistent with the disk fraction in our M4 galaxies derived from the S08 classification criteria and the visual classification schemes.

Nonetheless, part of our M3 and M4 galaxies that show disk-like kinematics may simply reflect the heterogeneous dynamical phases during the interactions \citep[e.g.][]{Mihos1998,Colina2005,Bellocchi2013}.
For instance, \citet{Bellocchi2013} show that only about half (6 out of 11) of their class 2 mergers (consistent with M3 and M4 in the definition of this paper) display complicated kinematics and the other half are classified as perturbed disks based on the visual classification scheme.
In this case, higher spatial resolution observations have the potential to resolve detailed kinematic information and trace the disturbance due to galaxy mergers.
Using the kinemetry-based classification scheme developed by S08, \citet{Goncalves2010} find that the merger fraction of a set of Lyman Break Analogs (LBAs) at $z\sim0.2$ decreases from $\sim70\%$ to $\sim38\%$ after redshifting their sample to $z=2.2$, which is partly due to the $\sim10$ times worse resolution in the redshifted datasets. 
IFS observations with high spatial resolution of local interacting systems will provide further insight on how spatial resolution have impact on the kinemetry-based analysis at different merger stages.

The significant overlap between single galaxies and interacting systems in the distribution of kinematic asymmetries (defined by simply averaging across all radius steps or weighted according to the circumferences) implies that a simple cut in $K_{asym}$ or $K_{asym,B12}$ is unable to differentiate mergers from isolated disks.
When the classification scheme in S08 is used, the success rate to classify S galaxies as disks is 100\%, yet the success rate to classify M3 galaxies as mergers is only 40\%.
The B12 scheme results in 100\% success rate in classifying M3 galaxies as mergers, whereas the success rate to classify S galaxies as disks decreases to only 50\%.
Meanwhile, we caution directly applying these merger recovery rate corrections to other work such as the samples in S08 or \citet{Alaghband-Zadeh2012} as this factor may vary depending on the physical properties of galaxy samples (e.g. merger stages, luminosity, mass, gas content).
Other indicators such as multi-wavelength morphological properties \citep[e.g.][]{Hammer2009} are necessary to further constrain the merger/disk classification of $z\sim1-3$ star-forming galaxies.

\section{Conclusions}
We have carried out merger/disk classifications for a set of 24 local (U)LIRGs based on the measurement of kinematic asymmetries and visual inspection.
Our sample spans a wide range of interaction sequence from single disk galaxies, interacting disks, to fully merged remnants.
To make a realistic comparison with the observations carried out for $z\sim1-3$ star-forming galaxies, we degrade the spatial resolution and sensitivity of the IFS datacubes as if the galaxies are observed at $z=1.5$ with a 0\arcsec.5 seeing.
The classifications of 28 local interacting systems and single disk galaxies lead to the following conclusions:

\begin{enumerate}

\item Based on the merger/disk fractions derived from kinematic classifications, we find that mergers clearly displaying two nuclei and tidal tails (M3 galaxies) show kinematic features most readily able to distinguish them from single disk galaxies. 
Nonetheless, some of our interacting disk pairs and merger remnants could be mistaken as disks when using kinemetry-based or visual classifications. 

\item We find that a high fraction of late-stage mergers (M4 galaxies) show signatures of rotating disks. This result may be explained as the disks reformed in the galaxy mergers or simply reflects the heterogeneous dynamical phases during the interactions.

\item Interacting disk pairs (M2 galaxies) that are resolved in the IFS observations have, as expected, smaller values of both $K_{asym}$ and $K_{asym,B12}$ compared to those pairs that are unresolved in the IFS observations.
Therefore, during the early interaction stages, kinematic classifications are only sensitive to the systems with small projected distances.

\item Our results show that kinematic properties alone are insufficient to constrain the merger fraction of $z\sim1-3$ star-forming galaxies.
Other merger indicators such as galaxy morphology traced by stars or molecular gas are required to further constrain the merger fraction at high$-z$.

\end{enumerate}

\acknowledgments
We thank the referee for the great effort with improving the paper.
C-LH thanks A. W. Mann for technical help with Figure~\ref{fig:wigs_single} \& \ref{fig:wigs_p1}.
C-LH wishes to acknowledge support from NASA grants NNX14AJ61G, NNX11AB02G, and the Smithsonian Predoctoral Fellowship program.
KLL and DBS gratefully acknowledge support from NASA grant NNX11AB02G.
CMC acknowledges support from a McCue Fellowship through the University of California, Irvine’s Center for Cosmology.
HAS acknowledges partial support from NASA grants NNX14AJ61G and NNX12AI55G.
CCH is grateful to the Klaus Tschira Foundation and Gordon and Betty Moore Foundation for financial support.

The Digitized Sky Surveys were produced at the Space Telescope Science Institute under U.S. Government grant NAG W-2166. The images of these surveys are based on photographic data obtained using the Oschin Schmidt Telescope on Palomar Mountain and the UK Schmidt Telescope. The plates were processed into the present compressed digital form with the permission of these institutions.

\appendix
\section{A. Optical Images, \Ha\ Velocity and Velocity Dispersion Maps of Redshifted WIGS Galaxies}

In this section, we present the optical images of the WIGS galaxies, velocity maps and velocity dispersion maps derived from the redshifted IFS datacubes.
We include all 28 galaxies that are used for kinematic classifications.

\begin{figure*}
\centering
  \includegraphics[width=0.85\textwidth]{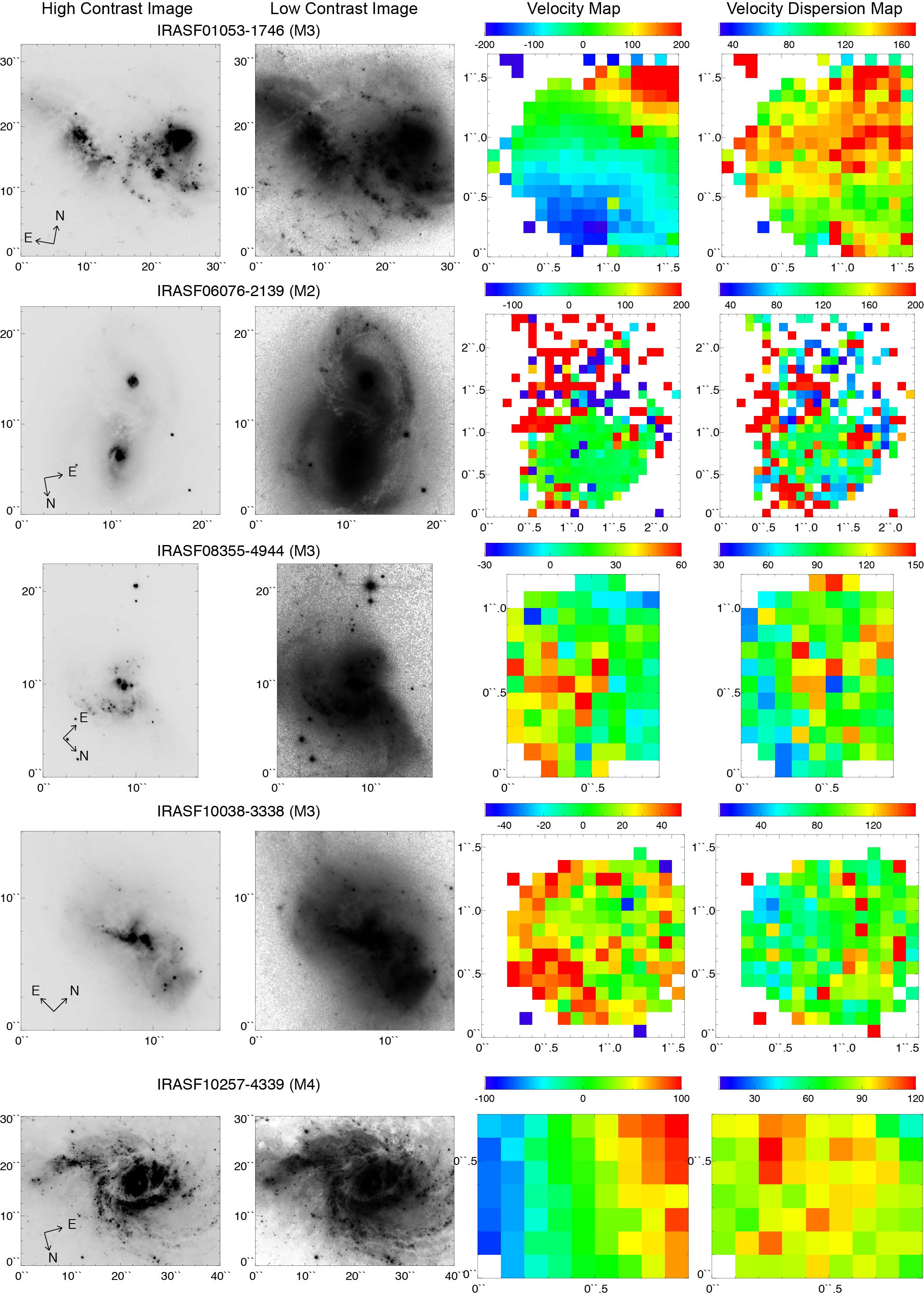} 
\caption{{\it Left and Left Middle}: Optical images and the morphological classifications of WIGS galaxies. 
The image types and filters are the same as those listed in Table~\ref{tab:data}. 
We present the optical images with two different stretching schemes (inverse hyperbolic sine and histogram equalization) that are optimized for demonstrating overall galaxy structure and revealing faint features.
{\it Right Middle and Right}: Velocity maps and velocity dispersion maps derived from the redshifted IFS datacubes.
These maps have the same field-of-view in physical scale (in parsec) as the optical images.
The angular sizes of the velocity and velocity dispersion maps are scaled as if the galaxies are observed at $z=1.5$.
The color bars are in the unit of km s$^{-1}$.
} 
\label{fig:wigs_p1}
\end{figure*}

\addtocounter{figure}{-1}
\begin{figure*}
\centering
  \includegraphics[width=0.85\textwidth]{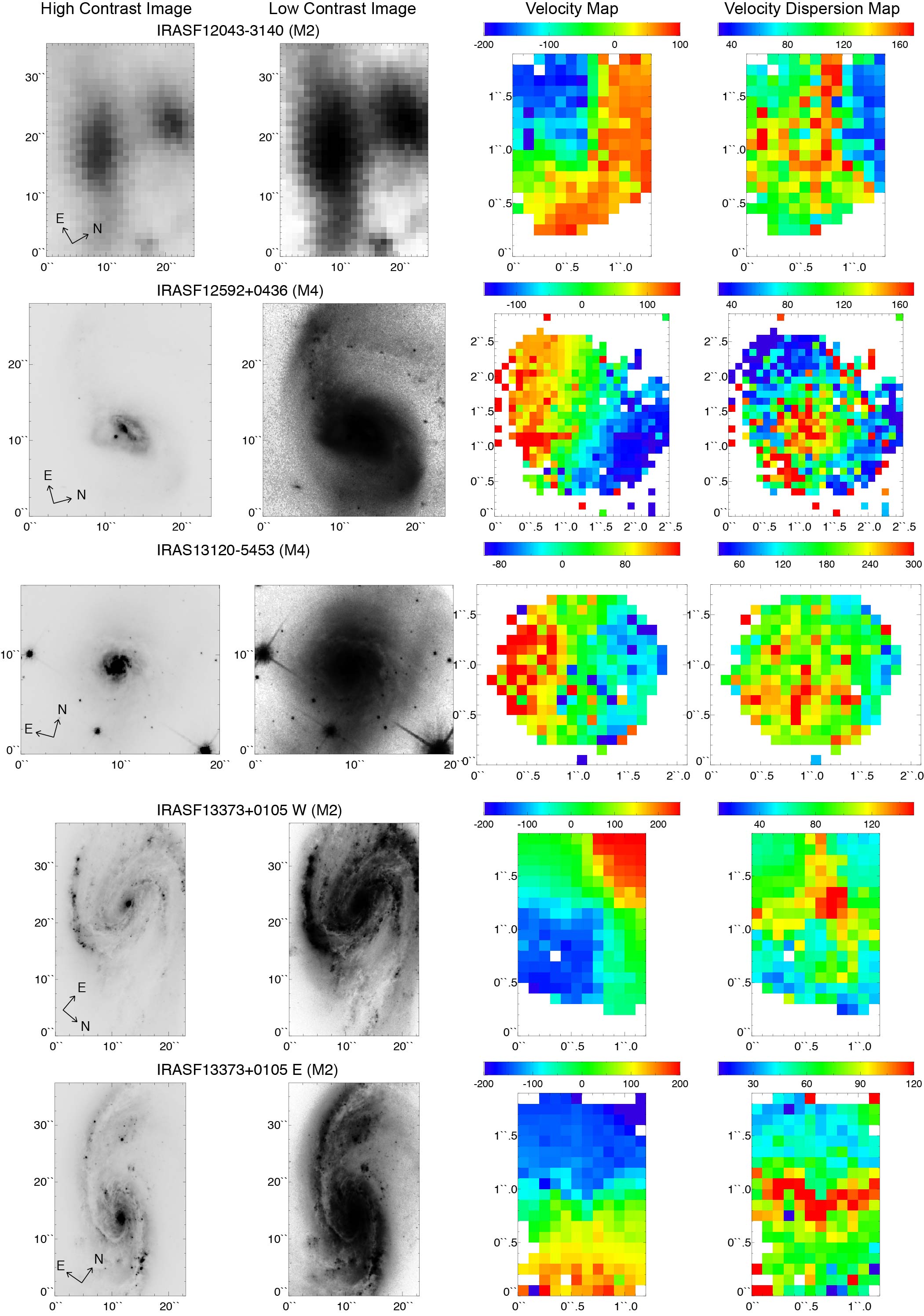} 
\caption{(Continued)
} 
\label{fig:wigs_p1}
\end{figure*}

\addtocounter{figure}{-1}
\begin{figure*}
\centering
  \includegraphics[width=0.85\textwidth]{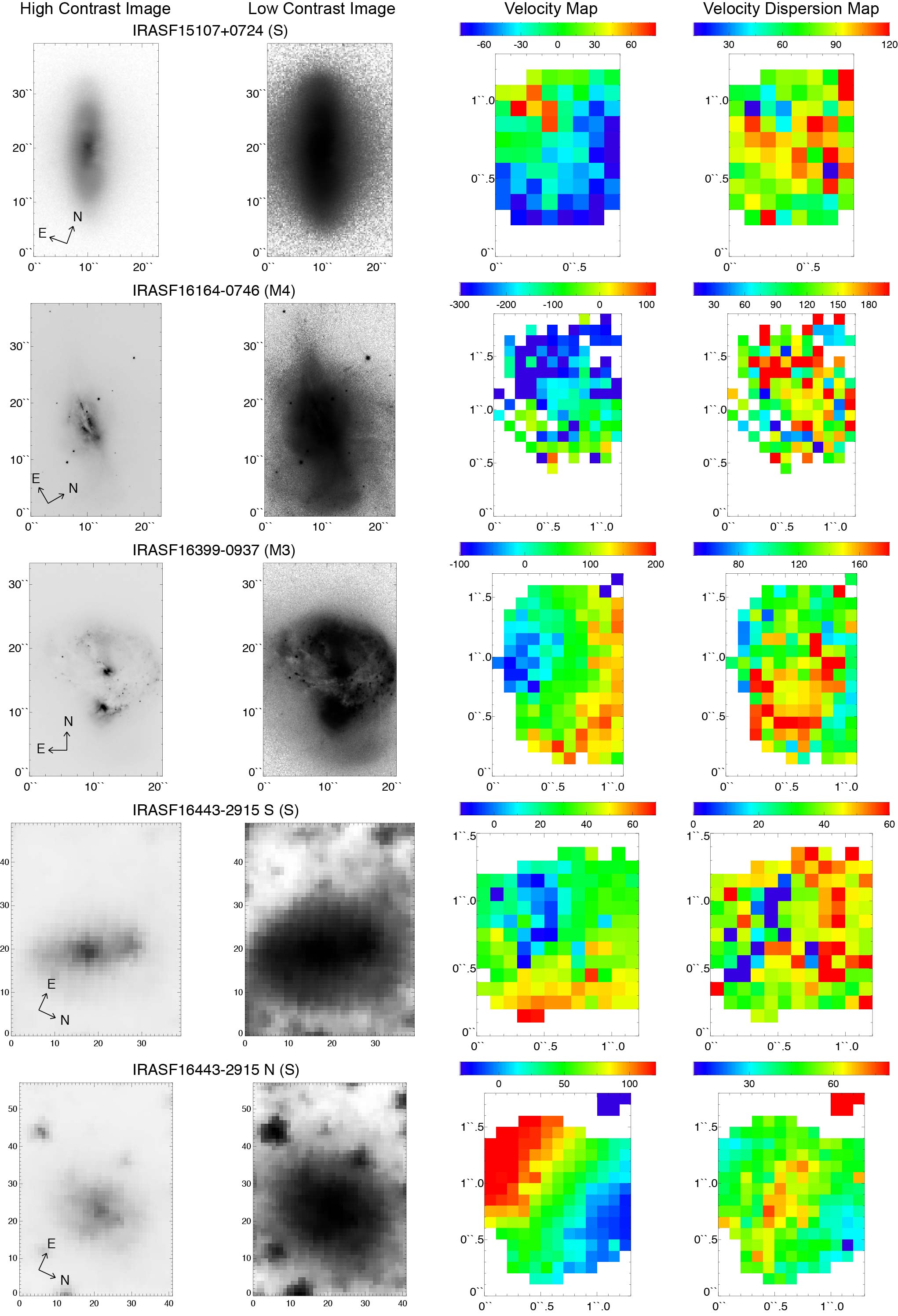} 
\caption{(Continued)
} 
\label{fig:wigs_p1}
\end{figure*}

\addtocounter{figure}{-1}
\begin{figure*}
\centering
  \includegraphics[width=0.85\textwidth]{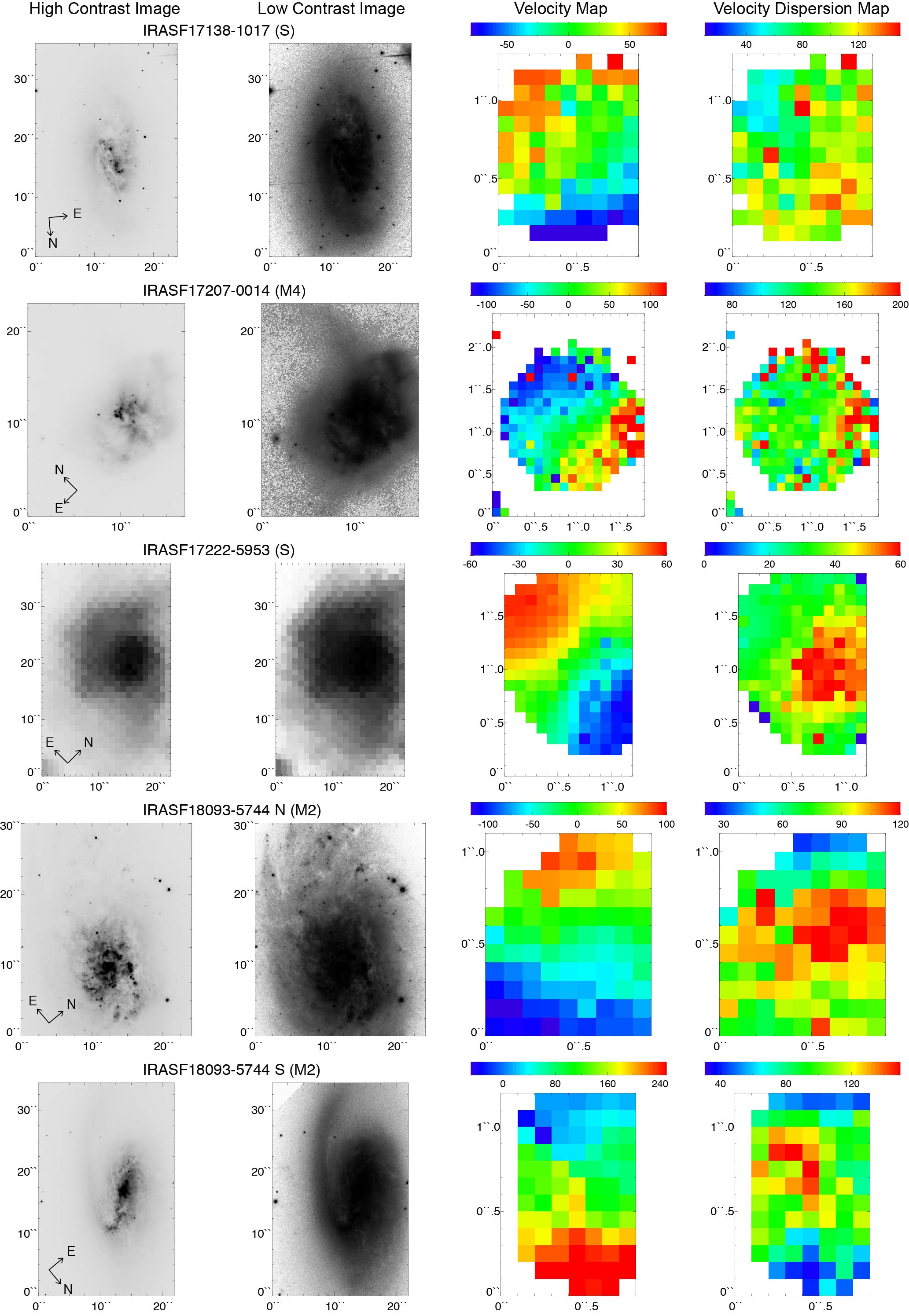} 
\caption{(Continued)
} 
\label{fig:wigs_p1}
\end{figure*}

\addtocounter{figure}{-1}
\begin{figure*}
\centering
  \includegraphics[width=0.85\textwidth]{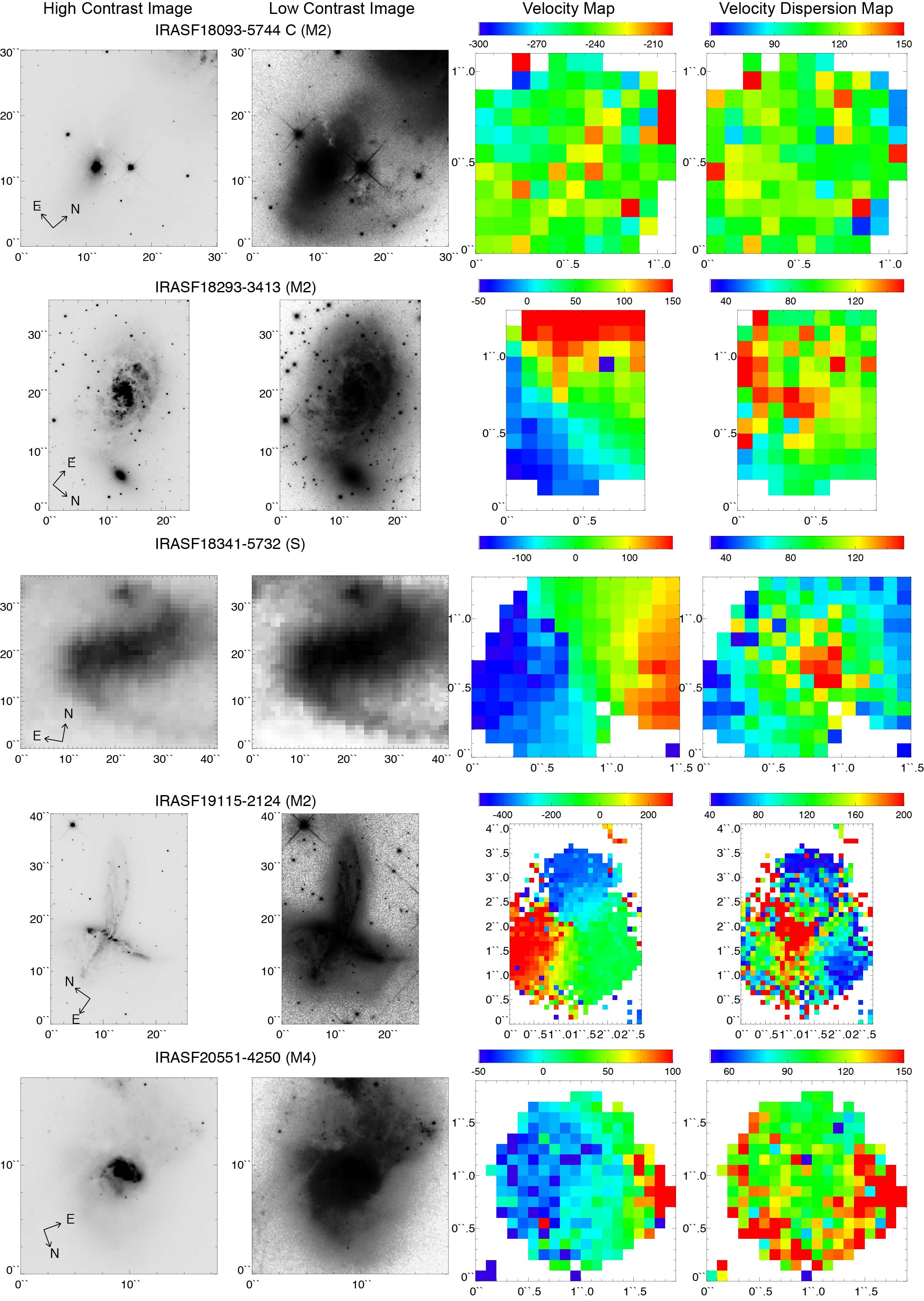} 
\caption{(Continued) We note that IRASF18093-5744 ``C'' refers to the central galaxy located in between the North and South galaxies.
} 
\label{fig:wigs_p1}
\end{figure*}

\addtocounter{figure}{-1}
\begin{figure*}
\centering
  \includegraphics[width=0.85\textwidth]{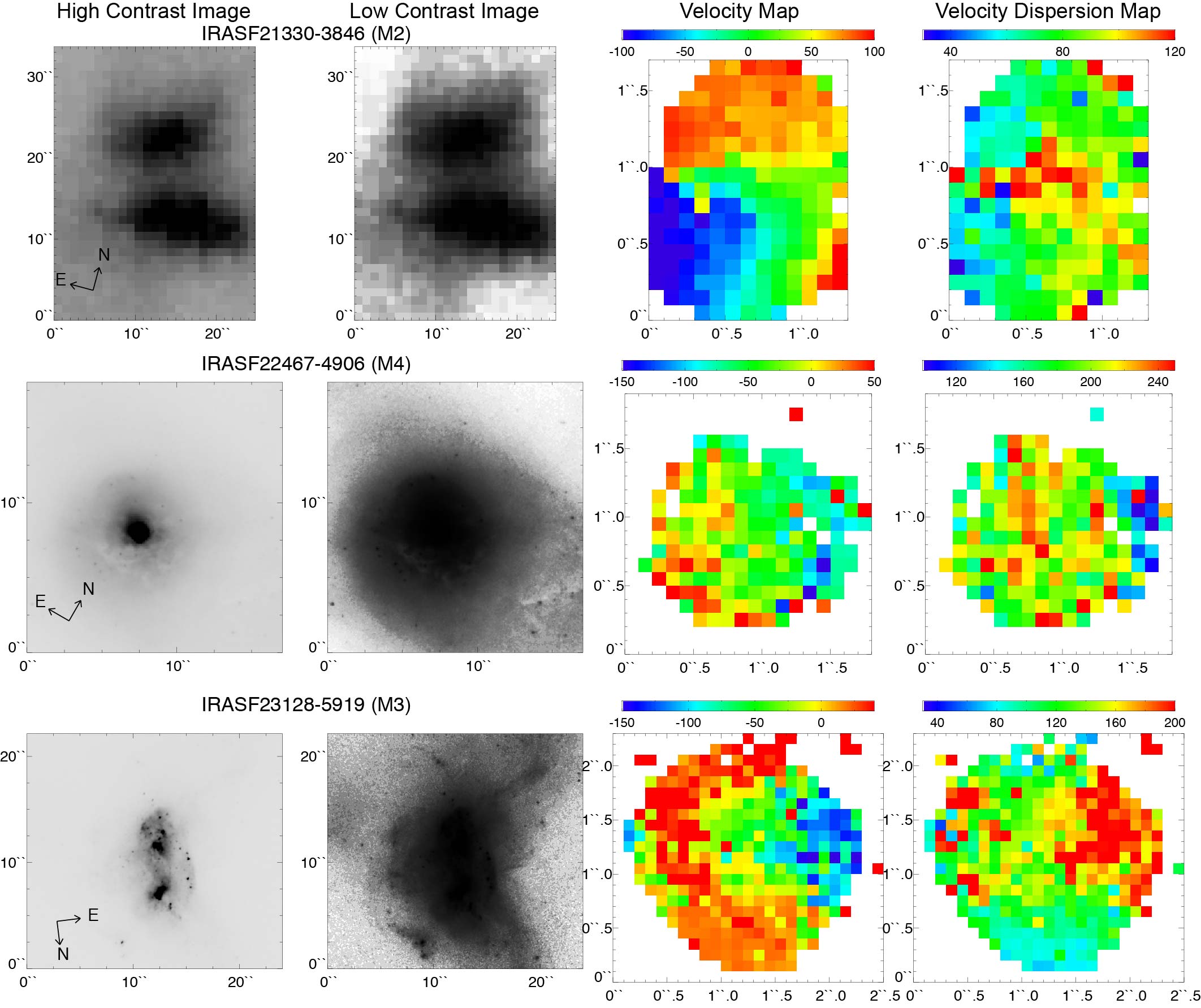} 
\caption{(Continued)
} 
\label{fig:wigs_p1}
\end{figure*}

\clearpage

\bibliographystyle{apj}
\bibliography{Cosmos}

\end{document}